\documentclass[aps,pre,twocolumn,showpacs,superscriptaddress,10pt,longbibliography]{revtex4-1}
\usepackage[T1]{fontenc}
\usepackage[dvipsnames,rgb,dvips]{xcolor}
\usepackage{float}
\usepackage{overpic}
\usepackage{bbm}
\usepackage{stmaryrd}
\usepackage{amsmath,amssymb,amsfonts}
\usepackage{hyperref}
\usepackage[normalem]{ulem}
\usepackage{mathrsfs}
\makeatletter
\def\amsbb{\use@mathgroup \M@U \symAMSb}
\makeatother
\usepackage[bbgreekl]{mathbbol}
\newcommand\ve[1]{\boldsymbol{#1}}
\newcommand{\ma}[1]{\ensuremath{\amsbb{#1}}}
\usepackage{graphicx}
\usepackage{bm}
\usepackage{subfig}
\usepackage{cleveref}
\usepackage{caption}
\usepackage{booktabs}
\newcommand{\TJ}{{T_{\rm J}}}
\newcommand{\lp}{\ensuremath{\left (}}
\newcommand{\rp}{\ensuremath{\right )}}
\newcommand{\lsp}{\ensuremath{\left [}}
\newcommand{\rsp}{\ensuremath{\right ]}}
\newcommand\minus{%
  \setbox0=\hbox{-}%
  \vcenter{%
    \hrule width\wd0 height \the\fontdimen8\textfont3%
  }%
}

\begin{document}

\title{Spinning and tumbling of micron-sized triangles in a micro-channel shear flow}

\author{J. Fries}
\email[]{These authors  contributed  equally to this work.}
\affiliation{Department of Physics, University of Gothenburg, SE-41296 Gothenburg, Sweden}
\author{M. Vijay Kumar}
\email[]{These authors  contributed  equally to this work.}
\affiliation{Department of Physics, University of Gothenburg, SE-41296 Gothenburg, Sweden}
\author{B. Mekonnen Mihiretie}
\affiliation{Department of Physics, University of Gothenburg, SE-41296 Gothenburg, Sweden}
\author{D. Hanstorp}
\affiliation{Department of Physics, University of Gothenburg, SE-41296 Gothenburg, Sweden}
\author{B. Mehlig}
\affiliation{Department of Physics, University of Gothenburg, SE-41296 Gothenburg, Sweden}
\begin{abstract}
We report on measurements of the angular dynamics of micron-sized equilaterally triangular platelets suspended in a micro-channel shear flow.  Our measurements confirm that such particles spin and tumble like a spheroid in a simple shear. Since the triangle has corners we can observe the spinning directly.  In general the spinning frequency is different from the tumbling frequency, and the spinning is affected by tumbling. This gives rise to doubly-periodic angular dynamics.
\end{abstract}
\maketitle

\section{Introduction}
The motion of very small particles in viscous flow is determined by two principles. 
First, the hydrodynamic force and torque on the particle are linearly related to its translational and rotational slip velocities, and to the strain-rate matrix of the flow. The constants of proportionality are the elements of the \lq resistance tensors\rq{} of the particle \cite{Happel:1983,Kim:2005}. These elements depend on the particle shape and on its orientation with respect to the flow. Second, the particle moves in such a way that the force and the torque on the particle vanish at every instant. 
So to compute the particle motion, one  must determine the resistance tensors by solving the Stokes problem with no-slip boundary conditions on the particle surface and the given flow velocities far from the particle.

For small neutrally buoyant spheroids this problem was solved by Jeffery \cite{Jeffery:1922}. He showed
that translation and angular dynamics decouple.
The centre-of-mass is advected by the flow.
In a simple shear flow the particle tumbles periodically. The vector $\ve n$ aligned with the axis of rotational symmetry of the spheroid follows one of infinitely many marginally stable periodic orbits, the \lq Jeffery orbits\rq{}.
More generally, symmetric 
particles with continuous rotation symmetry tumble in this way \cite{Bretherton:1962}.

\begin{figure}[t]
\begin{overpic}[width=0.4\columnwidth]{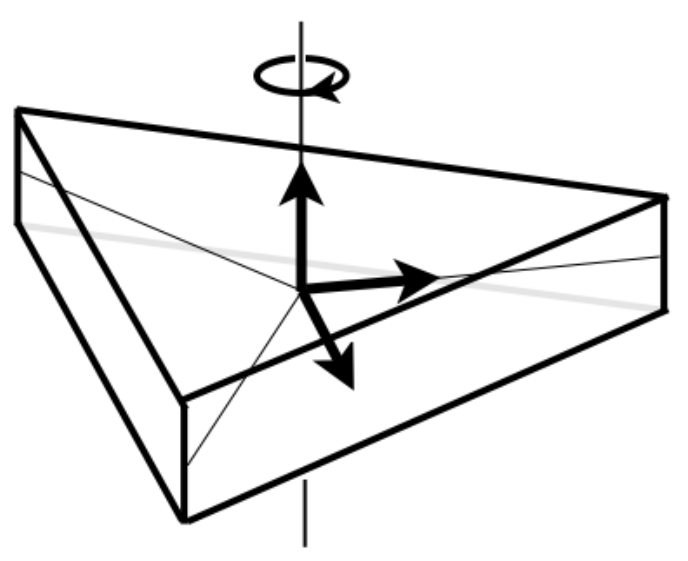}
\put(38,21){{$\ve q$}}
\put(60,30){{$\ve p$}}
\put(49,53){$\ve n$}
\end{overpic}
\caption{\label{fig:triangle} Triangular particle (schematic).  The unit vector $\ve n$ points along the axis of $3$-fold rotational symmetry, the unit vector $\ve p$ is orthogonal to $\ve n$ and points to a corner of the triangle, and $\ve q \equiv \ve p \wedge \ve n$.
}
\end{figure}

Often we encounter rigid particles that do not possess continuous
rotation symmetry.  Examples are hexagonal ice crystals \cite{Nak54}, 
and many plankton species \cite{Gua12}.
The algae {\em Triceratium}, for example
may assume the shape of flat triangles, squares, or pentagons \cite{AlgaeBase}.
In a recent paper \cite{angdyncry} we used symmetry considerations to analyse the
angular dynamics of such particles with discrete $k$-fold rotational symmetry 
($k > 2$) and certain mirror symmetries.
The theory predicts that the symmetry vector $\ve n$ follows a Jeffery orbit if the particle possesses a mirror plane that contains the axis of rotation.
An example of such a particle is the equilateral triangle shown schematically in Fig.~\ref{fig:triangle}.
Since this particle has corners, one can observe how the particle spins around $\ve n$.
This is described by the dynamics of the vector $\ve p$ that points to one of the corners of the triangle (Fig.~\ref{fig:triangle}).

To understand how small particles spin and tumble in turbulence is a question of current research \cite{Par12,Gus14,Ni14,Che13,Byron2015,Zha15,Voth16}.
In the past most experimental and numerical studies of the angular dynamics of small particles in viscous flow concentrated on non-spherical particles with a continuous rotation symmetry.
For such particles the tumbling of the symmetry vector $\ve n$ is readily observed.
The particle spin is however hard to measure experimentally, and it is often not analysed.
Exceptions are, first, some papers that analyse the angular dynamics of triaxial particles in a simple shear \cite{Hinch1979,Yarin1997,einarsson2016a}, using Jeffery's theory which
also applies to ellipsoids in its general form. Second, Voth {\em et al.} \cite{Mar14} measured the dynamics of small crosses in turbulence and showed that they tumble like spheroids.
Also in this case the spin can be determined.
Finally, Variano and co-workers studied the angular dynamics of agarose-hydrogel cylinders in turbulence,
and tracked particle spin by following micron-sized glass spheres contained within the cylinders using optical velocimetry 
\cite{Byron2015,Bor17}.  

\begin{figure}[t]
\begin{overpic}[width=0.95\columnwidth]{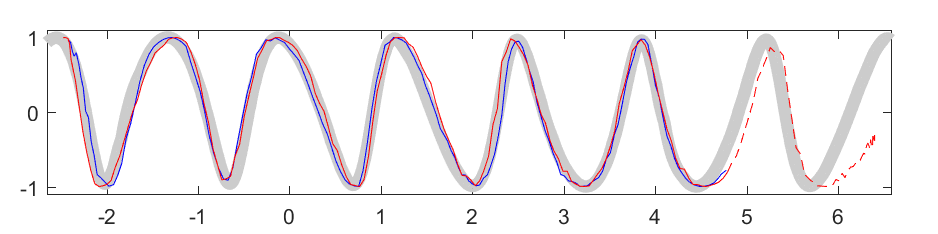}
\put(-2,17){$p_x$}
\end{overpic}
\begin{overpic}[width=0.95\columnwidth]{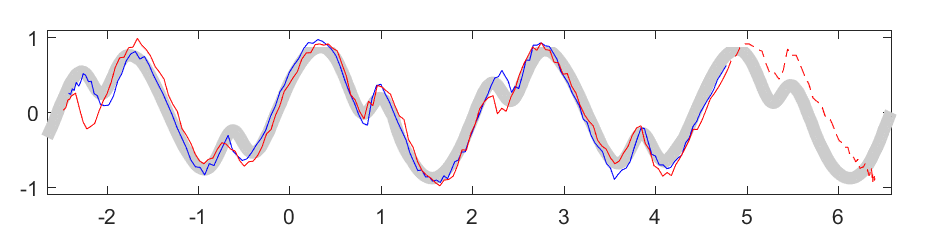}
\put(-2,17){$p_z$}
\end{overpic}
\begin{overpic}[width=0.95\columnwidth]{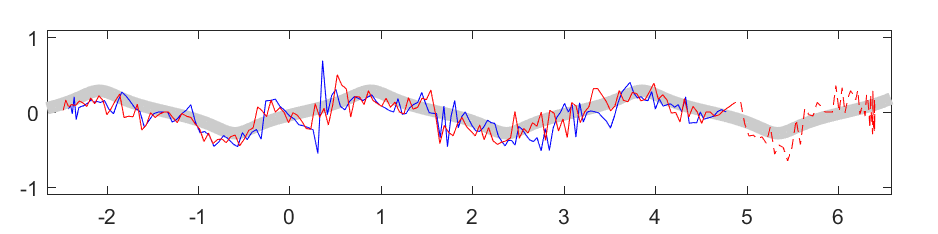}
\put(-2,17){$n_x$}
\end{overpic}
\begin{overpic}[width=0.95\columnwidth]{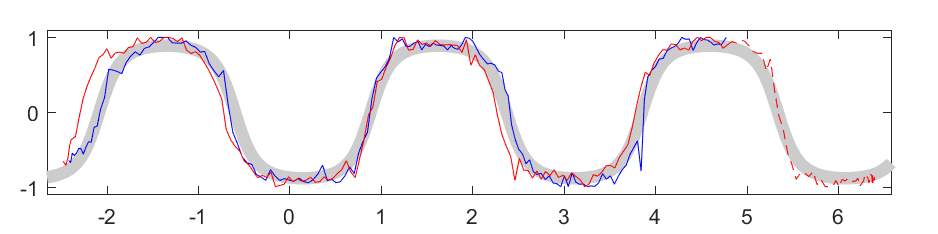}
\put(-2,17){$n_y$}
\end{overpic}
\begin{overpic}[width=0.95\columnwidth]{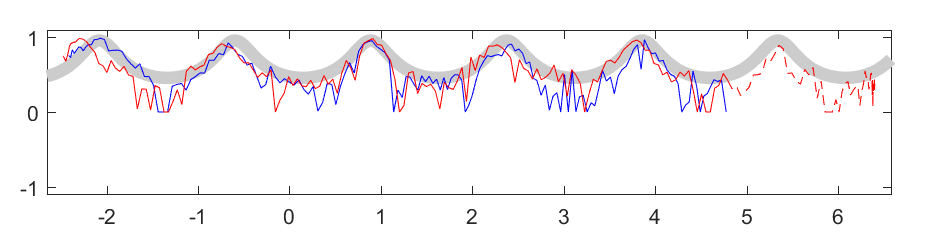}
\put(-2,17){$n_z$}
\put(53.,-2){$\Delta x$}
\end{overpic}
\caption{\label{fig:JO1} Angular dynamics of a triangular particle in a shear flow.
In the first two panels,  the $x$- and $z$-components of $\ve p$ (Fig.~\ref{fig:triangle}) are shown
as functions of the distance $\Delta x$ covered by the centre-of-mass of the particle in the channel direction.
Thin blue line: experimental data obtained before flow reversal. Thin red line: experimental data obtained after flow reversal.
Thick gray line: fit to Jeffery orbit as described in the text.
Dashed red and blue lines indicate data not part of a reversible
trajectory and therefore not fitted.
Panels~3 to 5 show the components of $\ve n$ computed from $p_x$ and $p_z$. Experimental parameters are shown in Table~\ref{tab:1} (trajectory~ID: 
A$_{\rm I}$).}
\end{figure}

\begin{figure}[t]
\begin{overpic}[width=0.95\columnwidth]{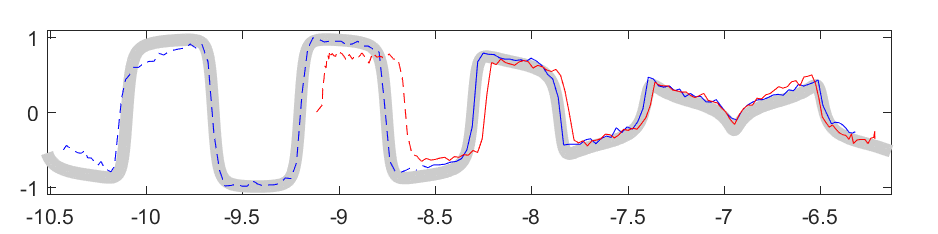}
\put(-2,17){$p_x$}
\end{overpic}
\begin{overpic}[width=0.95\columnwidth]{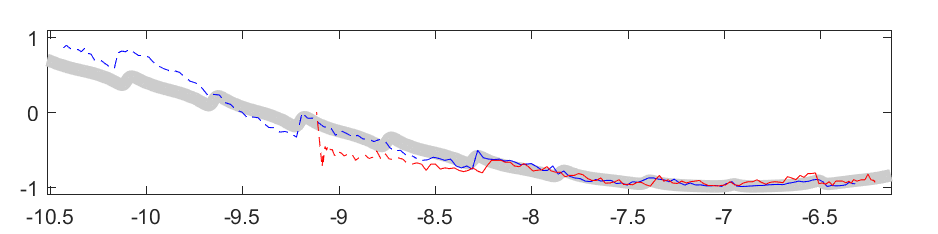}
\put(-2,17){$p_z$}
\end{overpic}
\begin{overpic}[width=0.95\columnwidth]{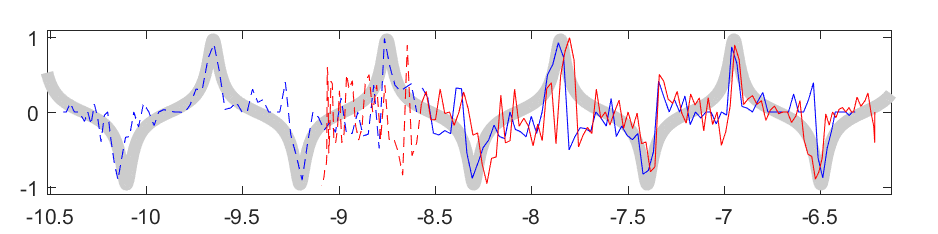}
\put(-2,17){$n_x$}
\end{overpic}
\begin{overpic}[width=0.95\columnwidth]{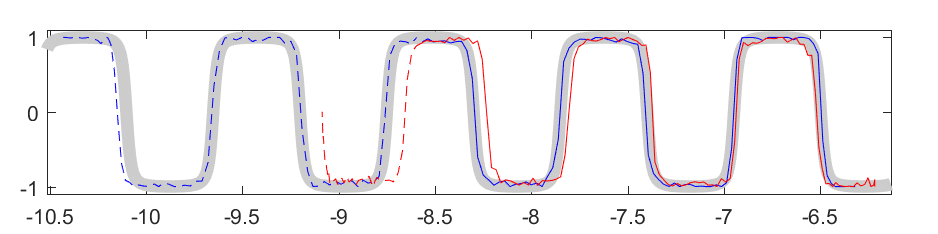}
\put(-2,17){$n_y$}
\end{overpic}
\begin{overpic}[width=0.95\columnwidth]{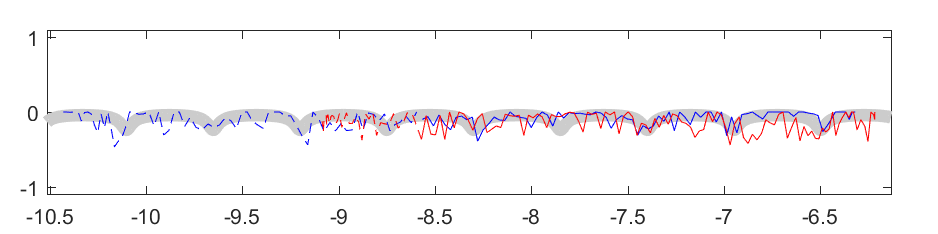}
\put(-2,17){$n_z$}
\put(49.,-2){$\Delta x$}
\end{overpic}
\caption{\label{fig:JO2} Same as Fig.~\ref{fig:JO1}, but for a different Jeffery orbit. Trajectory~ID: B$_{\rm I}$
(Table~\ref{tab:1}).}
\end{figure}

In this paper we report on measurements of the angular dynamics 
of neutrally buoyant micron-sized disk-like equilateral triangles that are suspended in
a micro-channel shear flow. These particles possess a $3$-fold rotation axis and a mirror plane that contains this axis.  Our measurements support the 
recent theoretical prediction \cite{angdyncry} that the angular dynamics of
such particles follows Jeffery orbits.
Figures \ref{fig:JO1}  and \ref{fig:JO2} show the dynamics of the vectors $\ve p$ and $\ve n$ corresponding to two different Jeffery orbits.
Also shown in these Figures are fits to Jeffery orbits (gray lines), in reasonable agreement with the experimental data.
We see that the dynamics of $\ve p$ is doubly periodic. This is explained by 
the fact that the dynamics of $\ve p$ (spinning) is coupled to the dynamics of $\ve n$ (tumbling), and that the two have different periods.
It is important that the angular dynamics is doubly periodic, because this implies that small shape changes that break the symmetry can result in chaotic motion. 
For nearly spheroidal particles this was shown theoretically \cite{Hinch1979,Yarin1997}. 
Ref.~\cite{einarsson2016a} verified the sensitivity of the angular dynamics to small shape changes experimentally
using glass rods with almost circular cross section in a micro-channel flow. 
For particles with discrete rotation symmetry this effect remains to be analysed. Here we concentrate on the doubly periodic angular dynamics.

The remainder of this paper is organised as follows. In Section \ref{sec:theory}
we briefly summarise the theoretical expectations. In Section \ref{sec:exp} we describe the experimental method. Section \ref{sec:results} contains our experimental results, their theoretical analysis, and a discussion of the experimental factors that we found difficult to control.
We conclude with Section \ref{sec:conclusions}. 
The Supplemental Material \cite{SM} contains additional plots of experimental data (similar to Figs.~\ref{fig:JO1} and \ref{fig:JO2})
as well as video recordings of the angular dynamics displayed in Figs.~\ref{fig:JO1} and \ref{fig:JO2}.

\section{Background and theory}
\label{sec:theory}
Figure~\ref{fig:triangle}
shows a platelet in the shape of an equilateral triangle (schematic). Its orientation is determined by the unit vectors $\ve n$ and $\ve p$, where $\ve n$ points along the axis of discrete rotation symmetry. The vector $\ve p$ is orthogonal to $\ve n$ and points to one of the corners of the triangle.
The basis is completed by $\ve q \equiv \ve p\wedge \ve n$. This coordinate system rotates with the particle.

Consider a particle suspended in a simple shear.
Define the lab-fixed coordinate system by the flow direction $\hat{\bf e}_x$, 
the shear direction $\hat{\bf e}_y$, and $\hat{\bf e}_z \equiv \hat {\bf e}_x \wedge \hat {\bf e}_y$, and denote 
the components of a vector $\ve v$ in this basis by $v_x$, $v_y$ and $v_z$.
Our experiments are conducted in a micro-channel flow.
Neutrally buoyant particles are placed in the channel cross section in such a way that the undisturbed fluid velocity near the particle takes the form $\ve U^\infty = \pm s y {\hat{\bf e}}_x$. 
Here $s >0$ denotes the undisturbed shear rate. 

We denote the symmetric part of the matrix of fluid-velocity gradients by $\ma S^\infty$,  the strain-rate matrix.
The asymmetric part, $\ma O^\infty$, describes the rotation of the flow.
Its elements are related to those of the vector $\ve \Omega^\infty \equiv \tfrac{1}{2} \ve \nabla \wedge
\ve U^\infty$ by $\ma O^\infty\ve x = \ve \Omega^\infty \wedge \ve x$ for an arbitrary vector $\ve x$.
The vector $\ve \Omega^\infty$ points in the  ${\mp} \hat{\bf e}_z$-direction (\lq vorticity direction\rq{}).

When diffusion and inertial effects are negligible then the centre-of-mass of triangle simply follows the flow.
Since the particle has a discrete threefold rotational symmetry, and since it has a mirror plane containing the axis of rotation, its angular dynamics 
is determined by Jeffery's angular velocity \cite{angdyncry}:
\begin{equation}
\label{eq:wp}
\ve \omega = \ve \Omega^\infty  -\Lambda \lp \ma S^\infty \ve n \rp \wedge \ve n\,.
\end{equation}
The corresponding
equations of motion for $\ve n$ and $\ve p$
read in vector notation \cite{einarsson2016a}:
\begin{subequations}
\label{eq:jeffery}
\begin{align}
\label{eq:n3}
\dot{\ve n} &= \ma O^\infty \ve n + \Lambda \lsp \ma S^\infty \ve n
- \lp \ve n\cdot \ma S^\infty \ve n\rp \ve n\rsp\,,\\
\label{eq:p}
\dot {\ve p} &= \ma O^\infty \ve p   - \Lambda \lp \ve n\cdot \ma S^\infty
\ve p\rp \ve n\,.
\end{align}
\end{subequations}
Dots denote time derivatives, and the \lq Bretherton constant\rq{} \cite{Bretherton:1962} $\Lambda$ parametrises the shape of the particle.
Oblate spheroids  have $-1 < \Lambda < 0$, and we expect that 
the shape parameter for our triangular platelets lies in this range. 
But we do not know the precise numerical value of $\Lambda$ for our particles. 
We remark that Eqs.~(\ref{eq:jeffery}) ensure that $\ve n$ and $\ve p$
remain orthonormal.

The dynamics of $\ve n$ is independent of that of $\ve p$. The vector $\ve n$ moves in one of infinitely many marginally stable Jeffery orbits, distinguished by the \lq orbit constant\rq{} $C$. Our definition of the orbit constant is slightly different 
from that used in Ref.~\cite{Hinch1979}, but equivalent. We use 
\begin{equation}
\label{eq:C} C \equiv n_z\quad\mbox{when}\quad n_y=0\,.
\end{equation}
When $C=0$, the vector $\ve n$ tumbles in the flow-shear plane. It aligns
for some time with the shear direction $\hat{\bf e}_y$ and then
rapidly flips by 180 degrees. By contrast, $C=\pm 1$ 
corresponds to the log-rolling orbit where $\ve n$ stays aligned with the vorticity direction.  
All Jeffery orbits have the same tumbling period $\TJ$ and frequency:
\begin{equation}
\label{eq:Tp}
\omega_{\rm T} \equiv\frac{2\pi}{\TJ}= \frac{s}{2}{\sqrt{1-\Lambda^2}}\,.
\end{equation}
Now consider the dynamics of $\ve p$. Eq.~(\ref{eq:wp}) shows that it depends on $\ve n$. The spinning frequency is given by 
\begin{equation}
\label{eq:wS}
\omega_{\rm S} \equiv \ve \omega \cdot \ve n =  \ve \Omega^\infty \cdot \ve n
\end{equation}
from Eq.~(\ref{eq:wp}).  The spinning frequency vanishes for tumbling in the flow-shear plane, and tends to $\pm$$s/2$ for log rolling.  
In general $\omega_{\rm S}$ varies as $\ve n$ tumbles, and it is in general different from $\omega_{\rm T}$.
Therefore the dynamics of $\ve p$ is doubly periodic. If we follow the $\ve p$-dynamics we can measure both $s$ and $\Lambda$.
Usually this is not possible because most experiments are conducted for particles with a continuous rotation symmetry.

Eqs.~(\ref{eq:jeffery}) require that rotational diffusion  \cite{Hinch1972} is negligible, which in turn requires that the P\'e{}clet number is large.
A lower bound for the P\'eclet number is given by ${\rm Pe} \equiv s/\mathscr{D}_{\rm max}$, where $\mathscr{D}_{\rm max}$ denotes the largest eigenvalue of the rotation diffusion tensor.  
For a flat disk with radius $a$ one estimates \cite{Kim:2005} 
$\mathscr{D}_{\rm max} \sim k_{\rm B} T/(8\pi \mu a^3)$,
independently of its thickness. 
Here $T$ is the temperature, $\mu$ is the dynamic viscosity of the fluid,
and $k_{\rm B}$ is the Boltzmann constant.
Also, Eqs.~(\ref{eq:jeffery}) 
are derived using the Stokes approximation, assuming that particle inertia \cite{einarsson2014} and fluid inertia  \cite{subramanian2005,einarsson2015a,einarsson2015b,candelier2015b,rosen2015d,Meibohm2016}  are negligible. For neutrally buoyant particles this is ensured by requiring that the shear Reynolds number ${\rm Re}_s = s a^2/\nu$ is small. Here $\nu$ is the kinematic viscosity of the fluid.

Stokes flow is reversible \cite{Taylor}. 
This ensures that Eqs.~(\ref{eq:jeffery}) too are reversible: the particle orientation retraces its angular dynamics upon reversal of the flow direction \cite{Bretherton:1962}.
Diffusion and inertial effects break reversibility.
Testing for reversibility allows us to rule out that diffusion and inertia have substantial effects, at least for the duration of the experiment.
\begin{figure}[t]
\begin{overpic}[width=0.9\columnwidth]{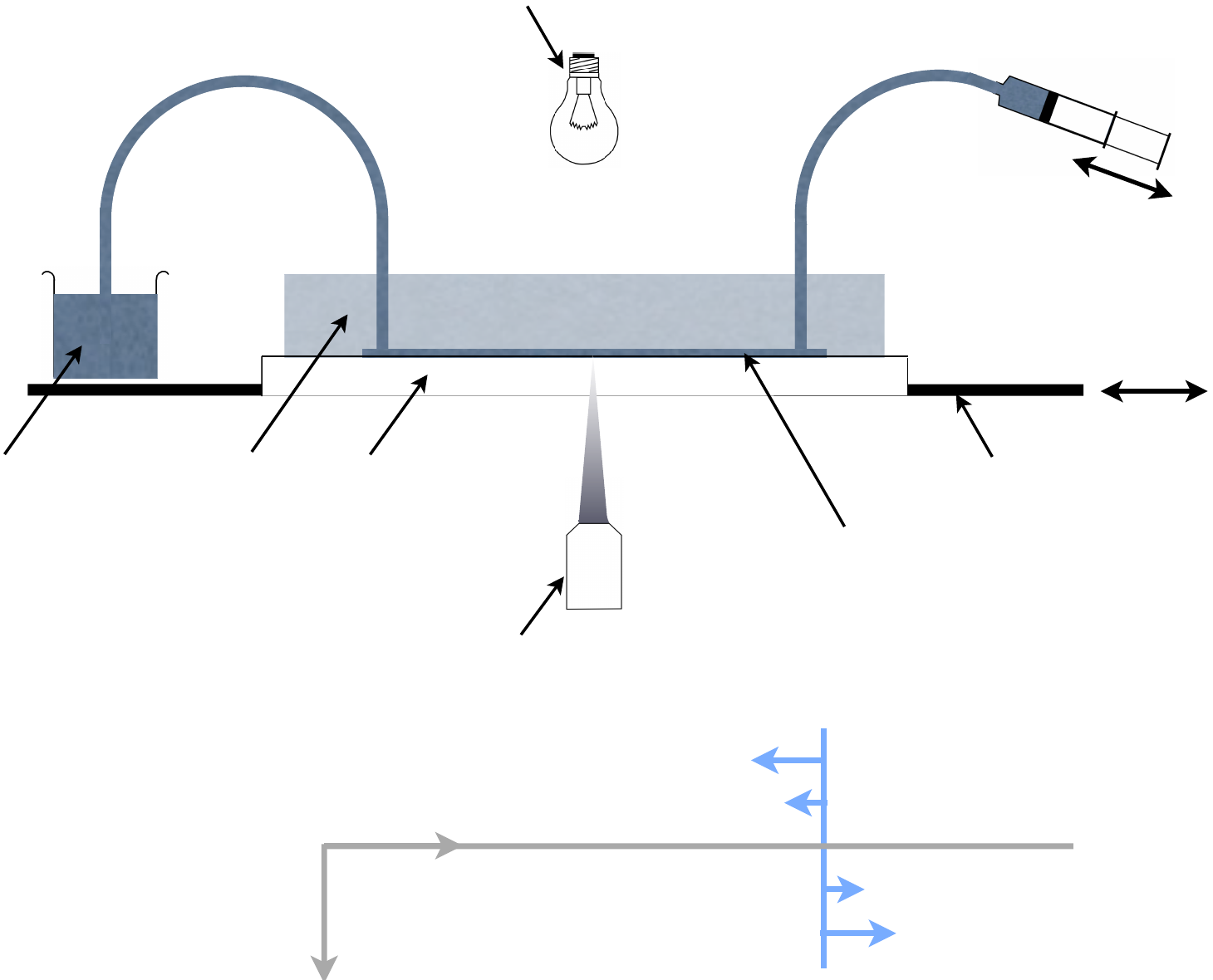}
\put(21,0){$\hat{\bf e}_y$}
\put(37,13){$\hat{\bf e}_x$}
\put(60.5,22){$U^\infty_x = s y$}
\put(32,25.1){microscope}
\put(-4,40){fluid}
\put(10,40){PDMS}
\put(73,40){moving stage}
\put(25,40){glass plate}
\put(59,34.5){micro-channel}
\put(90,61){pump}
\put(36,82){illumination}
\end{overpic}
\vspace*{5mm}
\caption{\label{fig:exp1} Experimental setup (schematic).
The $\hat{\bf e}_z$-axis points into the plane. Details are given in the text. 
}
\end{figure}
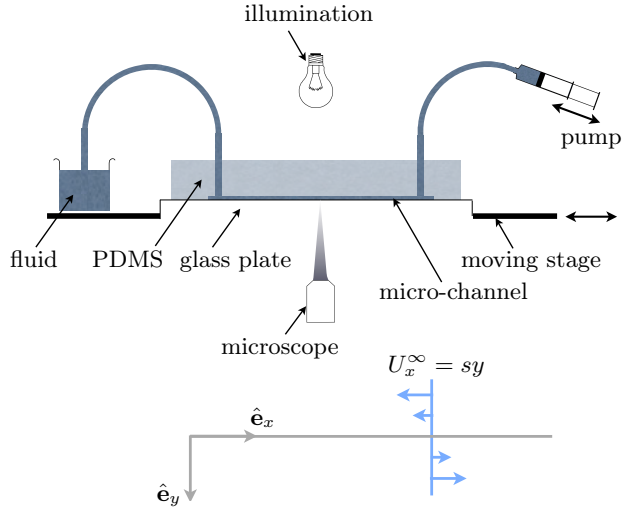

\section{Method}
\label{sec:exp}
The experimental setup is schematically shown in Fig.~\ref{fig:exp1}. It is similar 
to the setup used in Ref.~\cite{einarsson2016a}. The main differences are, firstly, that the particles
are different, and secondly that we used a different fluid in this paper, 
namely PEG-10K (aq.).

The density of the fluid was measured using an Anton Paar DMA 5000 instrument yielding a value of  $\rho$=1.05 g/cm$^3$.
Fluid-particle density matching was achieved by titration while observing the particle with the microscope.
The dynamic viscosity of the fluid was measured to 80.2 mPa$\cdot$s using an Ubbelohde viscometer at a temperature of 20 $^\circ$C.
During the viscosity measurement, the shear rate in tube of the viscometer 
was in the range 50 to 100 s$^{-1}$.
The density and viscosity measurements yield a kinematic viscosity of 76.3 mm$^2$/s.
Gonz\'a{}lez-Tello {\em et al.} \cite{Gonzalez:1994} have shown that similar polyethylen glycoles are Newtonian up to shear rates much higher than those used in this experiment.

Micro-channels were made out of polydimethylsiloxane (PDMS).
The channel was covered by a glass plate and fixed to a translation stage.
The channel dimensions were: length $40$(1) mm, width $2.5$(1) mm, and depth  $200$(10) $\,\mu$m.
In the experiment, the channel was mounted on a stage, illuminated from above, and imaged from below through the glass plate, using a Nikon X60 microscope objective.
The objective was fixed in the lab frame, while the stage was moved to keep the particle in the view of the objective.

For our measurements we considered only particles located 
near the horizontal channel centre, at approximately equal distance between
the channel walls. This was ensured by visual inspection of the particle
position in the camera plane, and it means that
the shear points in  the vertical direction (Fig.~\ref{fig:exp1}).
Our setup does not allow us to directly measure the depth (distance from the top or bottom plates) at which the particles are located. 
It is in principle  possible to estimate the particle depth by focusing on the particle with the microscope \cite{einarsson2016a}, 
but this method is not very precise. Therefore we decided to obtain the depth by fitting the angular particle dynamics
to Jeffery orbits in an assumed parabolic velocity profile.
\begin{figure}[t]
\begin{overpic}[width=0.95\columnwidth]{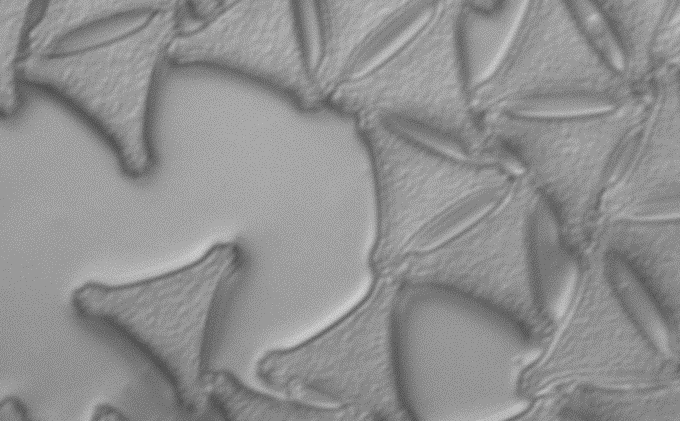}
\put(10,20){\rotatebox{17}{\color{white}\rule{2.1cm}{1pt}}}
\put(13,23){\rotatebox{17}{\textcolor{white}{\large $10\,\mu$m}}}
\end{overpic}
\caption{\label{fig:particles} Triangular particles obtained
from LIQUIDIA Technologies \cite{liquidia}. Reproduced with permission of LIQUIDIA Technologies. }
\end{figure}

We obtained $10\,\mu$m \lq pollen particles\rq{} from LIQUIDIA Technologies \cite{liquidia}. 
A representative image of the particles is shown in Fig.~\ref{fig:particles}. 
According to the data sheet provided by the company  \cite{liquidia}, the particles are triangular synthetic polymer micro-platelets, $0.5$ to $1\,\mu$m thick, and with side length $10\,\mu$m. The particle corners form an equilateral triangle with a center-to-corner distance $a = 5.8\,\mu$m. Their refractive index is between $1.5$ and $1.6$, and their mass density is $\rho_{\rm p} = 1.05\,$ g/cm$^3$.
Our particle suspension was highly diluted in order to avoid interaction between particles.

The flow rate in the experiments varied. The measurements reported in this paper
are consistent with a flow rate of the order of several $\mu {\rm l}\, {\rm min}^{-1}$.
Assuming a parabolic profile and a flow rate of 8 $\mu$l/min (see below), 
the flow velocity at the channel centre is $U_x^\infty=0.42$ mm/s.
A rough estimate (see below) says that the particle
was located between $20$ and $50\,\mu$m from the top or bottom plates. 
In a parabolic flow profile this means that the shear rate varied between 0 and $\sim$ 8 s$^{-1}$.
Together with the numbers quoted above this allows us to estimate the shear Reynolds number to ${\rm Re}_{\rm s}\equiv s a^2/\nu \sim 10^{-6}$, and the P\'e{}clet number to 
${\rm Pe}  \equiv s/ \mathscr{D}_{\rm max} \sim 10^{5}$.
This means that inertial effects are negligible on the time scales of the experiment, and that generally rotational diffusion is negligible.
When the particle aligns with the flow direction however, rotational diffusion may kick it out of alignment \cite{Hinch1972}.

To confirm that neither rotational diffusion nor inertial effects significantly affect the particle orientation during the experiment, we reverted the pressure and checked that the particle orientation retraces its trajectory.
In order to avoid large particle or fluid accelerations, the reversals were not instantaneous. 
The flow velocity was reduced slowly, reversed, and increased in a symmetric fashion.
Reversals lasted for approximately 10 {\rm s}.
During this time rotational diffusion and external noise may turn the particle and thus break reversibility.
As mentioned above, in this study we discarded data that did not exhibit reversal symmetry.

Since the flow speed $U^\infty_x$ varied during reversals, we plot the angular dynamics not as a function of time $t$ but as a function of the distance $\Delta x(t)$ that the centre-of-mass position of the particle covers: 
\begin{equation}\label{eq:deltax}
\Delta x(t) = \int_0^t\!\!{\rm d}t' \, U^\infty_x(t')\,.
\end{equation}
In the creeping-flow limit, changes in $U^\infty_x$ correspond to linear changes of the time scale
  \cite{Einarsson2013a}.  In this way we can disregard changes in $U^\infty_x$, provided that the Stokes approximation holds.
Reversibility implies that the angular dynamics collapses onto its reversed counterpart when plotted as a function of centre-of-mass position \cite{Bretherton:1962}.

The position of the translation stage was recorded at a rate of 33 Hz, and the video frames were recorded at a rate of 20 Hz.
We synchronised the two time series by visually matching two video frames with their respective stage positions.
The position of a video frame relative to the stage was then obtained by linear interpolation.
In any given video frame, the centre-of-mass position and the orientation
of the triangular particle were determined manually. Extracting the time series of $\ve p$ and $\ve n$ is complicated by two factors.
First, the $\ve n$-dynamics is sensitive to noise.
Second, the projection of a triangle onto the camera plane is ambiguous in that two differently oriented triangles have the same projection onto 
the camera plane. The two possible reconstructions are mirror images in the camera plane.
They have $p_x$, $p_z$ and $n_y$ in common, but $p_y$, and $n_x$ and $n_z$ are of opposite signs. 
Using that $\omega_z = n_x \dot{n}_y-n_y \dot{n}_x$ is parallel to the vorticity direction of the undisturbed flow, it follows that their motion corresponds to motion in opposite shears.
To resolve the ambiguity we chose the reconstruction for which $\omega_z \geq 0$ when $U_x^\infty > 0$.

Jeffery orbits were fitted to the particle trajectories measured before the reversal of the flow.
The fit minimizes the sum of squared distances between the experimental and theoretical
corner locations. The fitting parameters are $\Delta x(\TJ)$, $\Lambda$, $a$, and the three Euler angles describing the initial orientation
of the particle.
\begin{table*}
\caption{\label{tab:1} 
Summary of experimental data.
Steady flow speed $U^\infty_x$ estimated from centre-of-mass speed of particle.
Centre-of-mass distance $\Delta x(\TJ)$ covered in one Jeffery period $\TJ$, Bretherton constant $\Lambda$, orbit constant $C$ 
inferred from the fit to Jeffery orbit, shear rate $s$ inferred from $U^\infty_x$, $\Delta x(\TJ)$, and $\Lambda$
as described in the text.
Particle height above bottom inferred from $U^\infty_x$ and $s$.
Figures S1-S4 are in the Supplemental Material \cite{SM}. 
}
\begin{tabular}{llllllllll}
\hline\hline
Particle ID     & Trajectory ID  &Figure     & $U^\infty_x$  & $\Delta x(\TJ)$  & $\hspace*{3mm}\Lambda$\hspace*{3mm} & $C$ \hspace*{3mm} &  $s$ \hspace*{9mm} & height above   \\
           & &       & [mm/s]\,\,\,   &  [$\mu$m] &  &  &  [${\rm s}^{-1}$]  & bottom [$\mu$m]\,\,\,    \\[1mm]
\hline
A&I&(Fig.~\ref{fig:JO1}) \hspace*{3mm}& 0.31 & 2962 & -.95 &  $\, .96$ &  4.30 & 49\\
B&I&(Fig.~\ref{fig:JO2}) \hspace*{3mm}& -0.19 & -904 &  -.94 &   $\minus .25$ &  7.60 & 22 \\
A&II&(Fig.~S1) \hspace*{3mm}& 0.32 & 3097 &  -.96 &   -.93 &  4.47 & 49 \\
B&II& (Fig.~S2) \hspace*{3mm}& 0.17 & 811 &  -.95 &   -.72 &  8.50 & 18 \\
B&III&(Fig.~S3)  \hspace*{3mm}& 0.18 & 904 &  -.95 &   .71 &  8.23 & 19 \\
B&IV&(Fig.~S4) \hspace*{3mm}& 0.14 & 859 &  -.95 &  -.80 &  6.83 & 18 \\
\hline\hline
\end{tabular}
\end{table*}
\begin{figure}[b]
\begin{overpic}[width=0.9\columnwidth]{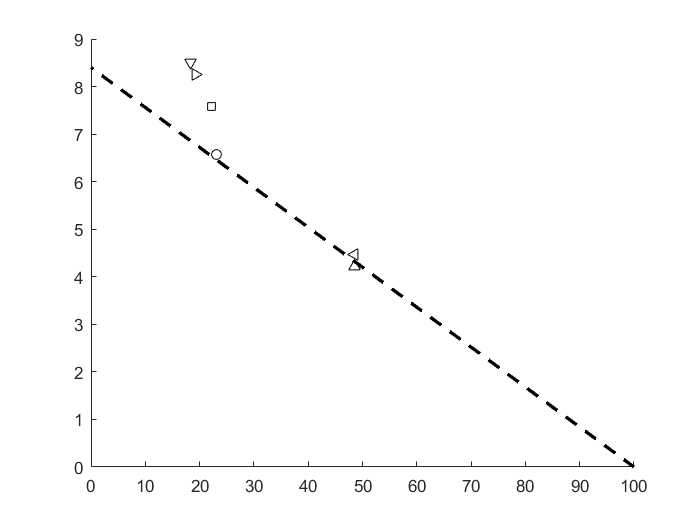}
\put(47,0){$y$ [$\mu$m]}
\put(3,35){\rotatebox{90}{$s$ [s$^{-1}$]}}
\end{overpic}
\begin{overpic}[width=0.9\columnwidth]{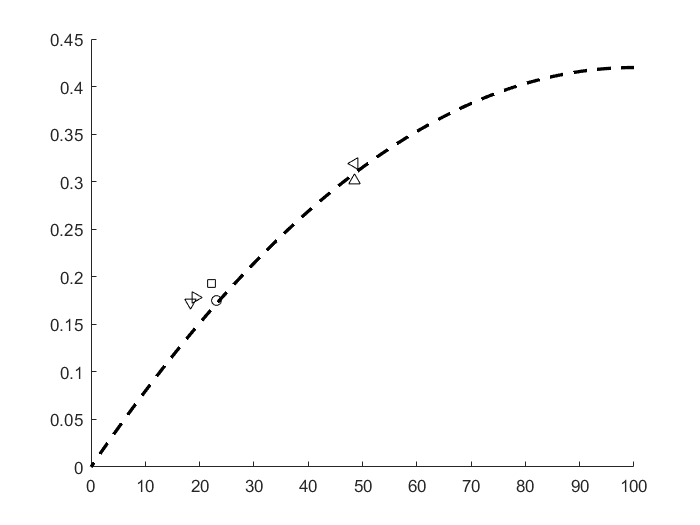}
\put(47,0){$y$ [$\mu$m]}
\put(1,30){\rotatebox{90}{$U_x^\infty$ [mm/s]}}
\end{overpic}
\caption{\label{fig:shear}
(Top) shear rate versus particle depth (distance from bottom plate),
as inferred from fits to Jeffery orbits (see text).
The dashed line is the shear rate at a flow rate of 8 $\mu$l/min.
Data are from the experiments summarised in Table~\ref{tab:1}: A$_{\rm I}$ ($\triangleleft$), A$_{\rm II}$ ($\vartriangle$), B$_{\rm I}$ ($\triangledown$),
B$_{\rm II}$ ($\triangleright$), B$_{\rm III}$ ($\oblong$),  and B$_{\rm IV}$ ($\circ$).
(Bottom) particle velocity vs. particle depth as inferred from fits to Jeffery orbits (see text).
The dashed line is the flow velocity at a flow rate of 8 $\mu$l/min, for a parabolic velocity profile.}
\end{figure}

\section{Results and discussion}
\label{sec:results}
The first two panels of Fig.~\ref{fig:JO1} show how the $x$- and $z$-components of the vector $\ve p$ change as the particle travels along the channel. We show two trajectories, measured before and after reverting the channel flow (thin blue and red lines). 
We see that the dynamics reverses fairly well, so that inertial effects and rotational diffusion are negligible. In steady flow this is expected,
since ${\rm Re}_s$ is small and ${\rm Pe}$ is large. But it is important to note that
we estimated the P\'e{}clet number for steady flow. During reversals the shear rate $s$ tends to zero.
Reversals last for approximately 10 {\rm s}, and during this time rotational diffusion and external noise can turn the particle and thus break reversibility. In this study we discarded data that does not exhibit reversal symmetry.

The gray lines in Fig.~\ref{fig:JO1} show fits of Eqs.~(\ref{eq:jeffery}) to the experimental $\ve p$-dynamics.
From the fits we can infer $s$, $\Lambda$, and $C$.  The values obtained are shown in Table~\ref{tab:1}. 
The fit results in $\Lambda \approx -0.95$. For the orbit constant we find $C\approx{0.96}$, quite close to unity. This means that the angular dynamics is close to the log-rolling orbit where the triangle lies mostly flat in the flow-shear plane. The corresponding video-microscopy recording is included in the Supplemental Material \cite{SM}. 
The $\ve n$-dynamics is shown in the lower panels of Fig.~\ref{fig:JO1}. The data is noisier than the $\ve p$-dynamics because
it is reconstructed from a projection of the $\ve p$-dynamics.
But it is clear that $n_z$ is always quite close to unity, as it must for a log-rolling orbit. 

We measured the centre-of-mass velocity of the particle, a measure of the undisturbed fluid velocity $U_x^\infty$ at the particle position.
We obtained $U_x^\infty \sim 0.31$mm/s.
The fluctuations in $U_x^\infty$ were negligible for large parts of the particle trajectory.
Using $\Delta x(\TJ) = U_x^\infty \TJ$ we obtained the Jeffery period from the measured value of $\Delta x(\TJ)$.
Using Eq.~(\ref{eq:Tp}) and the fitted value of $\Lambda$ we determined the shear rate to $s \approx 4.3 \,${\rm s}$^{-1}$.
These values of $U_x^\infty$ and $s$ are consistent with a flow rate  $8 \,\mu {\rm l}\, {\rm min}^{-1}$ if one assumes a parabolic flow profile, and that the particle was located at a distance of $50\,\mu$ from the bottom or
top plates.  This flow rate and particle depth yield $U^\infty_x = 0.31\, {\rm mm}\, {\rm s}^{-1}$ and $s = 4.2 \,{\rm s}^{-1}$ (dashed lines in Fig.~\ref{fig:shear}).

As pointed out in the Introduction, we can determine both tumbling and spinning frequencies by fitting the experimental data to Jeffery orbits.
For Fig.~\ref{fig:JO1}, the tumbling frequency (\ref{eq:Tp}) evaluates to $\omega_{\rm T} = 0.65\,${\rm s}$^{-1}$.
The spinning frequency, $\omega_{\rm S}$, changes between
$1.0\,${\rm s}$^{-1}$ and $2.1\,${\rm s}$^{-1}$ for the smallest and largest values of $n_z$.
Since the $\ve p$-dynamics couples to that of $\ve n$, and since the spinning and tumbling frequencies
are different, the $\ve p$-dynamics is doubly periodic (first two panels of  Fig.~\ref{fig:JO1}). The $\ve n$-dynamics, by contrast, is approximately periodic - disregarding the noise in the experimental data.

Fig.~\ref{fig:JO2} shows data for a different Jeffery orbit.
The corresponding values of $\Lambda$, $C$, and $s$ are given in Table~\ref{tab:1}.
A slight dephasing of an otherwise reversible motion is seen in this experiment.
This may be due to a small density mismatch between particle and fluid that causes the particle to rise or sink.
For this experiment, we find $\omega_{\rm T} = 1.35 \,${\rm s}$^{-1}$, and $\omega_{\rm S}$ changes between $0.18\,${\rm s}$^{-1}$ and $0.94\,${\rm s}$^{-1}$.
The orbit constant for this particle is $C = -0.25$, so the $\ve n$-vector tumbles near the flow-shear plane.

More experimental results are given in the Supplemental Material \cite{SM}, Figs.~S1-S4.
All data described in the main text and in the Supplemental Material was obtained
for two particles, labeled 'A' and 'B' in Table~\ref{tab:1}.

The inferred shear rates, particle velocities, and particle depths for all experiments are also given in this Table.
Fig.~\ref{fig:shear} shows how the shear rate $s$ and velocity $U_x^\infty$ depend on the particle depth, as obtained
from the fits to Jeffery orbits (symbols).
Also shown are the expected dependencies assuming a parabolic flow profile with a flow rate of  8 $\mu$l/min (dashed lines).
Most data points display reasonable agreement with this flow rate.
The fits imply that particle A was located approximately $\sim$ 50 $\mu$m from the top or bottom plates, and particle B was located at $\sim$ 20 $\mu$m.
In the latter case, the theory based on a flow rate of 8 $\mu$l/min underestimates the measured shear rates somewhat. 

We remark that Fig.~S2 shows that particle B moves reversibly, but deviates somewhat from the Jeffery orbit fitted.
We see that the $\ve n$-dynamics agrees well with the fit, but that there are deviations in the $\ve p$-dynamics.
Since the wall distance for this particle was only about 20 $\mu$m, these deviations could be caused by particle-wall interactions.
Numerical simulations \cite{Mody:2005} show that the $\ve n$-dynamics of an oblate spheroid tumbling close to a flat wall is slightly modified by the presence of the wall. The trajectory in Fig.~S2 may indicate that the particle spin is more sensitive to the presence of the wall than the tumbling.

\begin{figure}[t]
\begin{overpic}[width=0.9\columnwidth]{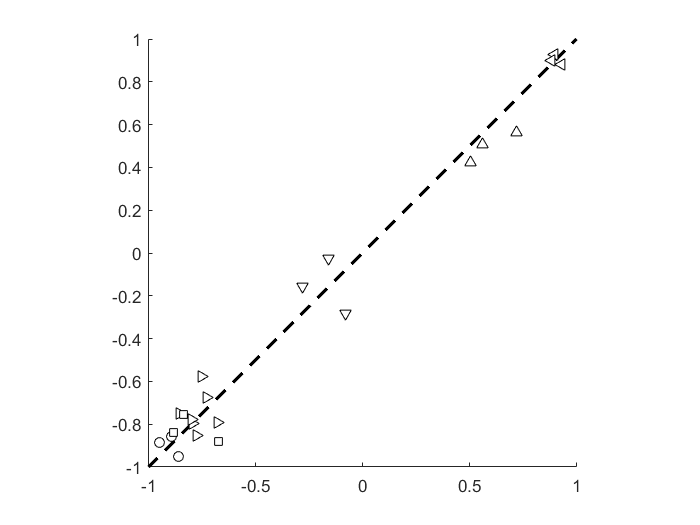}
\put(50,0){$n_z^{(i)}$}
\put(10,35){\rotatebox{90}{$n_z^{(i+1)}$}}
\end{overpic}
\caption{\label{fig:nz} Periodic dynamics of $n_z$.
We measured $n_z$ when $n_y=0$
to obtain a discrete
time series $n_z^{(i)}$ for $i = 1,2,\ldots$. The graph shows
$n_z^{(i+1)}$ versus $n_z^{(i)}$. Symbols as in Fig.~\ref{fig:shear}. Also shown is the theoretical expectation (dashed line) derived from
the periodicity of the $\ve n$-dynamics: $n_z^{(i+1)} = n_z^{(i)}$. }
\end{figure}

Finally, Fig.~\ref{fig:nz} shows a summary of the experimentally determined periodic $\ve n$-dynamics.
The Figure shows subsequent values of $n_z$ at $n_y=0$, for the orbits shown in Figs.~\ref{fig:JO1}, \ref{fig:JO2}, and S1-S4. 
Since the $\ve n$ dynamics is periodic, these values of $n_z$ should remain constant, equal to the orbit constant $C$ \cite{einarsson2016a},
as shown by the dashed line in Fig.~\ref{fig:nz}. We see that the experimental data is roughly consistent with this expectation.
In the experiment the values fluctuate (Fig.~\ref{fig:nz}).
There are several sources of error that could explain the scatter.
First, inaccuracies in the manual determination of the particle corners in the experimental images could be caused by systematic changes in the diffraction patterns upon changing orientation.
Second, effects of particle-wall interactions cannot be excluded, especially since some particles were located near the wall.  
Wall irregularities may enhance the effect.
Third, we cannot rule out that small particle-shape irregularities may cause deviations from the dynamics
predicted for symmetric particles \cite{einarsson2016a}. We note 
that Figs.~\ref{fig:JO1}, \ref{fig:JO2}, and S1-S4 show reversible angular dynamics. This rules out that the fluctuations are due to rotational diffusion.

\section{Conclusions}
\label{sec:conclusions}
We measured the angular dynamics of micron-sized platelets with $3$-fold rotation symmetry in a micro-channel shear flow.
The angular dynamics is well described by Jeffery's equation of motion.  
This supports the conclusion \cite{angdyncry} that a particle with a discrete $k$-fold rotation symmetry ($k>2$) and a mirror plane containing the rotation axis moves according to Jeffery's equation. The resulting angular trajectories are doubly periodic
(Figs.~\ref{fig:JO1} and \ref{fig:JO2}) because tumbling and spinning frequencies are not commensurate.

Our measurements are not  accurate enough to systematically study the effect of minute shape asymmetries. But since we can track the corners of the particle, we could in principle describe the particle orientation by its Euler 
angles \cite{Goldstein} $(\theta,\phi,\psi)$ and represent the particle dynamics by a Poincar\'e{} surface-of-section \cite{Yarin1997,einarsson2016a}. This would make it possible to study how breaking of the discrete rotation symmetry may give rise to chaotic tumbling.
It would also be interesting to measure the angular dynamics of particles with
$k$-fold rotation symmetry that lacks a mirror plane containing the rotation axis. We predicted \cite{angdyncry} that there are such particles that tumble like spheroids, but spins differently: $\ve \omega\cdot\ve n$ does not vanish when $\ve n$ tumbles in the flow-shear plane.

\acknowledgements{We thank E. Variano and G. Voth for discussions. We thank LIQUIDIA for permission to reproduce Fig.~\ref{fig:particles}. We acknowledge support by Vetenskapsr\aa{}det [grant numbers 2013-3992 and 2017-03865], Formas [grant number 2014-585],  by the grant \lq Bottlenecks for particle growth in turbulent aerosols\rq{} from the Knut and Alice Wallenberg Foundation, Dnr. KAW 2014.0048, 
by the Carl Trygger Foundation for Scientific Research, and by the MPNS COST Action MP1305 \lq Flowing matter\rq{}.}


%
\end{document}